\documentclass[twocolumn,showpacs,prb,amsmath,amssymb]{revtex4}

\usepackage{graphicx}
\usepackage{dcolumn}
\usepackage{bm}

\begin{document}

\preprint{APS/123-QED}

\title{Ac Susceptibility and Static Magnetization Measurements of CeRu$_2$Si$_2$ at Small Magnetic Fields and Ultra Low Temperatures}

\author{D. Takahashi$^1$, S. Abe$^1$, H. Mizuno$^1$, D. A. Tayurskii$^{1,2}$, K. Matsumoto$^1$, H. Suzuki$^1$ and Y. \={O}nuki$^3$}
\affiliation{
$^{1}$Department of Physics, Kanazawa University, Kakuma-machi, Kanazawa 920-1192, Japan\\
$^{2}$Physics Department, Kazan State University, Kremlevskaya str., 18, Kazan, 420008, Russia\\
$^{3}$Graduate School of Science, Osaka University, Toyonaka, Osaka 560-0043, Japan}

\date{\today}

\begin{abstract}
The magnetic properties of CeRu$_2$Si$_2$ at microkelvin
temperatures (down to 170 $\mu$K) and ultra small magnetic fields
($0.02\!\sim\!6.21$ mT) are investigated experimentally for the first
time. The simultaneously measured ac susceptibility and static
magnetization show neither evidence of the magnetic
ordering, superconductivity down to the lowest temperatures nor
conventional Landau Fermi-Liquid behavior. The results imply the
magnetic transition temperature in undoped CeRu$_2$Si$_2$ is very close
to absolute 0 K. The possibility for proximity of CeRu$_2$Si$_2$ to
the quantum critical point without any doping is discussed.

\end{abstract}

\maketitle

The unusual properties of heavy fermion (HF) systems are determined by the competition between intersite spin couplings,
Ruderman-Kittel-Kasuya-Yosida interaction, and intrasite Kondo interaction.~\cite{Doniach}
In a system dominated by the Kondo effect, the Pauli paramagnetic (PP) state with massive quasiparticles
is achieved through screening of the \emph{f} electron's magnetic moments by conduction electrons
below the characteristic temperature $T_\text{K}$.
The physical properties of the HF compounds below $T_\text{K}$ are well understood within
the framework of the Landau Fermi-liquid (LFL) theory.

Recently, however, non-Fermi-liquid (NFL) behavior was observed in a large class of HF compounds
near the quantum critical point (QCP).~\cite{QCP1,QCP2}
NFL systems exhibit anomalous temperature dependence of the physical quantities in contrast to the LFL theory,
such as specific heat $\Delta C/T \!\! \propto \! -\ln T$,
resistivity $\Delta \rho \! \propto \! T ^{\epsilon} (1\!\leq\!\epsilon\!<\!2)$
and magnetic susceptibility $\Delta \chi \! \propto$ either $1- \! \sqrt{T}$ or $\!- \ln T$.
In general, the quantum (zero-temperature) phase transition is driven by a control parameter
other than temperature, for example, composition, pressure, or magnetic field,
and is accompanied by a qualitative change in the correlations in the ground state.
The second order quantum phase transitions and QCPs in HF systems can
be classified into two types: (i) the long-wavelength fluctuations of the order parameter
are the only critical degrees of freedom and the quantum criticality is developed as
spin-density wave instability,~\cite{SCRHF, Millis} here the zero-temperature spin fluctuations
are given by the Gaussian fluctuations of the order parameter;
(ii) local critical modes coexist with long-wavelength fluctuations of the order parameter
and there is non-Gaussian distribution of the fluctuations.~\cite{Si}
These are the so-called locally critical phase transitions where
the quantum criticality of CeCu$_{(6-x)}$Au$_x$ (Ref.~\onlinecite{CeCu6Au}) and YbRh$_2$Si$_2$ (Ref.~\onlinecite{YbRhSi1})
are regarded as type-(ii) QCP.~\cite{Si}

CeRu$_2$Si$_2$ with a ThCr$_2$Si$_2$-type crystal structure is well known to be a typical HF
compound with an electronic specific-heat coefficient $\gamma\!=\!350 \text{mJ/K}^2\text{mol}$
below $T_\text{K}\!=\!20$ K.~\cite{Gamma, KondoTemp}
This compound exhibits the pseudo-metamagnetic transition into the ferromagnetically ordered state induced by
the magnetic field at $H_\text{M}\!=\!7.8$ T below 10 K.~\cite{meta1,meta2,meta3,meta4,meta5}
The neutron-scattering measurements note short-range antiferromagnetic (AFM) correlations in
CeRu$_2$Si$_2$ even below $T_\text{K}$.
These time-fluctuating correlations are described by different incommensurate wave vectors.~\cite{AFM1,AFM2}
A $\mu$SR experiment shows an ultra small static moment of the order of 10$^{-3}\mu_\text{B}$.~\cite{mSR}
It is most remarkable that the alloying compound systems Ce$_{(1-x)}$La$_{x}$Ru$_2$Si$_2$ and
Ce(Ru$_{(1-x)}$Rh$_x$)$_2$Si$_2$ show an incommensurate spin density wave (SDW) ground state
in a concentration range $x\!>\!0.08$ (Ref.~\onlinecite{La1}) and $0.03\!<\!x\!<\!0.4$ (Ref.~\onlinecite{Rh1}).
This long-range ordering has the form of a sine-wave modulated structure with the short-range
correlation of CeRu$_2$Si$_2$, as described above.
At the critical concentrations of $x_{c}\!=\!0.075$ for La and $x_{c}\!=\!0.03$ for Rh doping,
the SDW transition vanishes but $T_\text{K}$ remains a finite temperature.
Therefore, these small critical concentrations of La and Rh suggest that CeRu$_2$Si$_2$
might be located in the vicinity of the type-(i) QCP, and AFM spin fluctuations are expected
to play a key role in this magnetic ground state.
The $C$, $\chi$ and $\rho$ measurements show the conventional LFL ground state for CeRu$_2$Si$_2$
below $T_\text{K}$ down to 20 mK.~\cite{meta1}
All magnetization and susceptibility measurements have been performed at the magnetic field above 1 T and
at low temperatures.
In this letter, we report for the first time the results of ac susceptibility and static
magnetization measurements at microkelvin temperatures (down to 170 $\mu$K) and
ultra small magnetic fields ($0.02\!\sim\!6.21$ mT).
The obtained magnetic field and temperature dependence of the susceptibility and magnetization provide
evidence of  NFL behavior and allow us to think about the proximity of CeRu$_2$Si$_2$ to QCP.

The single crystal of CeRu$_2$Si$_2$ was prepared by a Czochralski pulling method with
starting materials Ce ($99.99$\%), Si ($>\!99.999$\%) and Ru ($99.99$\%) and purified by a solid state
transport method.
The sample size was $11\times4.2\times1.5$ mm$^{3}$.
The sample was cooled with a copper nuclear demagnetization refrigerator and a $^3$He-$^4$He dilution refrigerator.
It  was sandwiched between two silver plates which were parts of the thermal link to the copper nuclear stage.
The temperature was measured by a Pt NMR thermometer, a $^3$He melting
curve thermometer~\cite{MeltingCurve} and a RuO$_2$ resistance thermometer.
All these thermometers and the thermal link were attached to the same flange of the nuclear stage.
The temperature difference between sample and heat bath was estimated to be less than the
order of 0.1 \% at all temperatures.

The ac susceptibility and static magnetization of CeRu$_2$Si$_2$ were measured simultaneously in a static field
$0.02\!\leq \!B\! \leq \!6.21$ mT by an ac impedance bridge using a SQUID magnetometer.
The applied static field declined a few tens of degrees from the crystalline \emph{c}-axis.
All of the ac susceptibility measurements were performed at a frequency of 16 Hz with an excitation field below
$0.75$ $\mu$T parallel to the static field.
The primary coil, secondary coil, and static field coil were placed inside a Nb superconducting magnetic shield,
which was surrounded by a $\mu$ metal magnetic shield to suppress any external stray field.
The static magnetization was calibrated against the absolute value measured by another magnetometer
in the temperature range from 4 to 2 K.

\begin{figure}[htb]
\includegraphics[width=7cm,keepaspectratio]{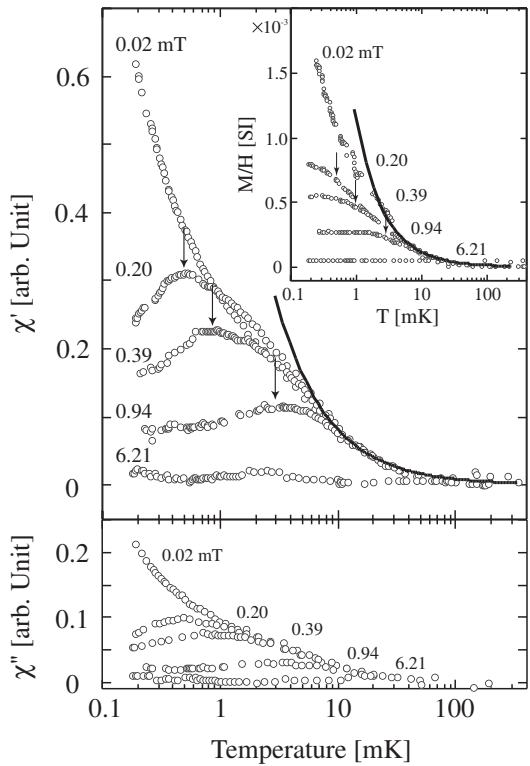}
\caption{\label{ACDC} Temperature dependence (on a logarithmic
scale) of the ac susceptibility ($\partial M/ \partial H$) at
different applied fields as indicated in the figure.
The inset shows the static susceptibility ($M/H$) in the same temperature range.
The arrows and solid line indicate the peak temperature observed by ac susceptibility and the Curie
law at each figure, respectively.}
\end{figure}
The ac susceptibility was measured during cooling and warming, and the results showed no appreciable
hysteresis.
The static magnetization was measured in the warming procedure.
Figure \ref{ACDC} shows the temperature dependence of the inphase ($\chi'$) and quadrature ($\chi''$) components
of the ac susceptibility ($\partial M/\partial H$) at different magnetic fields below 400 mK.
The inset of Fig.\ref{ACDC} shows the temperature dependence of the static susceptibility ($M/H$)
derived from the static magnetization.
We calibrated all data against the temperature independent PP susceptibility which observed above $\sim$50 mK.
Below $\sim$50 mK, we observed an excess susceptibility obeying the Curie law.

The ac susceptibility shows a peak at the magnetic field between 0.20 and 0.94 mT.
The peak temperatures $T_\text{P}$ shift to higher values and the ac susceptibility is suppressed with
increase in the applied magnetic field.
In particular, the ac susceptibility at 6.21 mT is suppressed almost to the level of the PP susceptibility.
The ac and static susceptibilities deviate from the Curie law as they approach $T_\text{P}$.
The Curie constant $C$ can be written in the form $C \!=\! N_\text{A} \mu _\text{0} \mu _\text{p} ^2 / 3 k _\text{B} V _\text{mol}$.
The effective magnetic moment $\mu_\text{p}$ turns out to be $0.020\,\pm\,0.003 \mu _\text{B}$/unit cell
from the static susceptibility and is independent of applied magnetic fields.
The value of $\mu_\text{p}$ is in agreement with the ultra-small static moment observed in the $\mu$SR experiment.~\cite{mSR}
The static susceptibility, however, becomes flat with no peaks in fields higher than 0.20 mT.
\begin{figure}[tpbh]
\includegraphics[width=7cm,keepaspectratio]{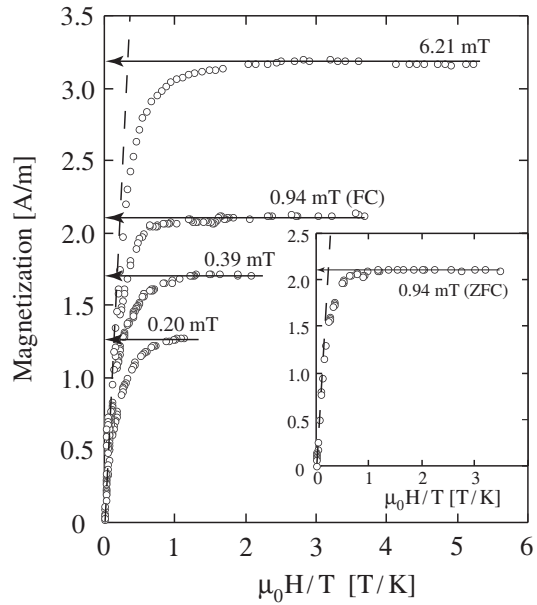}
\caption{\label{H/T} $H/T$ ($H$ is the applied field) dependence of
the static magnetization of CeRu$_2$Si$_2$ above 0.20 mT at ultra
low temperatures. The inset shows the $H/T$ dependence of
magnetization at 0.94 mT obtained in the zero field cooling
experiment. Solid lines (saturated magnetization) and dashed lines
(Curie-law) are  guides for the eye.}
\end{figure}

Figure \ref{H/T} shows the $H/T$ dependence of the static magnetization below 400 mK.
The dashed line corresponds to the Curie law with $\mu_\text{p} = 0.02 \mu_\text{B}$.
In the fields above 0.20 mT, the magnetization clearly shows the saturation.
The saturated magnetic moment $\mu_\text{s}$ can be evaluated in each field using the following
relation : $M_\text{s} \!=\! N_\text{A} \mu_\text{s} / V_\text{mol}$.
The calculated $\mu_\text{s}$ are $1.20 \times 10^{-5}, 1.60 \times10^{-5}, 1.95 \times 10^{-5} $
and $2.98 \times 10^{-5} \mu_\text{B}$/unit cell, and the ratio of $\mu_\text{p}$ to $\mu_\text{s}$ is derived
as $1.80 \times 10^{3}, 1.15 \times 10^{3}, 0.98 \times 10^{3}$ and $0.73 \times 10^{3}$
at 0.20, 0.39, 0.94 and 6.21 mT, respectively.
Figure \ref{H/T} suggests that the magnetization cannot be attributed to an impurity effect.
At very small concentrations, the magnetic impurities contribution to the total magnetization should
behave according to a Brillouin function.
With increasing concentration, a locally ordered state like spin-glass can be formed.
The impurity effect on the static magnetization in CeCu$_\text{6}$ at low temperatures is one example for the first case.~\cite{CeCu6static}
It indicates that the ratio of $\mu_\text{p}$, as deduced by Curie law, to $\mu_\text{s}$ has to be of the order of 1.
However, this does not agree with our results, $\mu_\text{p}/\mu_\text{s} \!\sim\! 10^3$.
In the second case, many compounds with a spin glass transition show quite different
magnetization behavior between zero field cooling (ZFC) and field cooling (FC) measurements through the transition temperature.
The inset in Fig. \ref{H/T} shows the magnetization at 0.94 mT measured by ZFC.
The results of ZFC and FC do not indicate different behavior below $T_\text{P}$ in this field.
Consequently, the possibility of spin glass transition is also strongly denied.

\begin{figure}
\includegraphics[width=8cm,keepaspectratio]{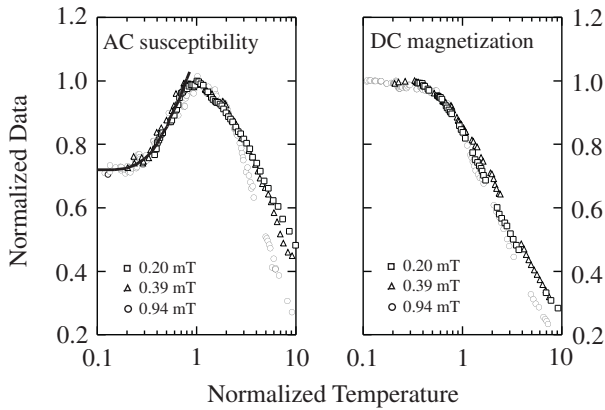}
\caption{\label{Scaling}Scaling behavior of the ac susceptibility
and static magnetization in the fields between 0.20 and 0.94 mT.
The vertical axes represent the normalized susceptibility and the normalized magnetization.
The horizontal axis is the normalized temperature.
Details of normalization and the solid line are explained in the text.}
\end{figure}
Below $T_\text{P}$, there is unique scaling behavior of the ac susceptibility and static magnetization
from 0.20 to 0.94 mT, as shown in Fig. \ref{Scaling}.
We normalized the susceptibility by its peak height and the magnetization by the saturated value at each field.
And the temperature is normalized by $T_\text{P}$.
The scaling behavior provides evidence of the proximity of CeRu$_2$Si$_2$ at small magnetic fields and ultra low
temperatures to some phase transition.~\cite{S-K.Ma}
The nature of such transition is determined by interplaying between ferromagnetic (FM) and AFM fluctuations
observed by neutron scattering experiments.~\cite{meta4,meta5,AFM1,AFM2}
And this type of interplaying was evidently observed around $H_\text{M}$.~\cite{meta4}
In order to shed light on the nature of this transition we analyzed this scaling behavior from
the viewpoint of up-to-date theories for ordering in HF compounds.

According to the mean-field theory, the temperature dependence of the susceptibility with the SDW ground state below
N\'{e}el temperature obeys the following expression : $\chi(T)\!=\!\chi_{0} + B \exp(-a/T)$, where $\chi_{0}$ and $a$ are
the $\chi(T \!\rightarrow\! 0)$ and an energy gap, respectively.
There, the ratio of the gap energy to $T_\text{N}$ should be above 1.76.
In our case, the normalized temperature $T^*$ dependence of the normalized ac susceptibility $\chi^{*}(T^{*})$ is
obeying above the expression with $a/T^{*} \!\approx\! 1.0$ at the peak temperature, as shown in Fig. \ref{Scaling}.
This ratio is in contrast to the SDW state case and there is no indication for the AFM transition at $T_\text{P}$.
The scaling behavior and the exponent type temperature dependence of $\chi^{*}$ below $T_\text{P}$, however,
suggest that CeRu$_2$Si$_2$ is in some magnetic field arranged state between 0.20 and 0.94 mT.
We speculate that the physical background of this scaling behavior is the quantum critical fluctuation effect of
CeRu$_2$Si$_2$ which is in proximity to the QCP discussed below. 
 
Further, we compare CeRu$_2$Si$_2$ with CeCu$_6$ which is also a typical HF compound
and very similar to CeRu$_2$Si$_2$ in its 4\emph{f} electron behavior below $T_\text{K}$.
Recently, it has been shown that magnetic ground state of CeCu$_6$ is the SDW state and 
$T_\text{N}$ is in fair agreement with the estimated one by the self consistent renormalization
(SCR) spin fluctuation theory.~\cite{Tsujii}
The SCR theory for the HF system predicts the value of $T_\text{N}$ as an equation : $T_\text{N} \!=\! 0.1376 p_\text{Q}
^{4/3} T_A ^{2/3} T_0 ^{1/3}$, where $p_\text{Q}$ is the staggered spontaneous moment in $\mu_{B}$ at $T = 0$ K,
$T_\text{A}$ and $T_\text{0}$ are the characteristic temperatures
in the \textbf{\emph{q}} and \textbf{$\omega$} space, respectively.~\cite{SCRHF}
If we use the derived values of  $T_\text{A} \!=\! 16$ K, $T_\text{0} \!=\! 14.1$ K (Ref.~\onlinecite{CharacteristicTemp})
and $p_\text{Q} \!= \!7 \!\times\! 10^{-3}$ $\mu_{B}$ (Ref.~\onlinecite{mSR}) for CeRu$_2$Si$_2$,
the predicted $T_\text{N}$ is estimated as $T_\text{N} \sim 2.8$ mK.
However, the magnetic properties of CeRu$_2$Si$_2$ do not show any indication of the magnetic ordering
in the smallest applied magnetic field.
The estimation of $T_\text{N}$ for SDW state in framework of SCR theory takes into account only AFM characteristic wave vectors
while our data indicate an existence of both FM and AFM fluctuations.
And these two types of fluctuations can lead to some disordered state in the smallest field.
This means that the magnetic transition temperature in CeRu$_2$Si$_2$ is possibly close to $T = 0$ K and
the spin system of CeRu$_2$Si$_2$ under our conditions is in the vicinity of the QCP.

In the case of type-(i) QCP, we cannot explain the temperature dependence of the magnetic susceptibility in CeRu$_2$Si$_2$
only from AFM fluctuations which have been observed in neutron experiments.~\cite{AFM1,AFM2}
Based on the SCR theory, the uniform susceptibility of an itinerant weak AFM compound should not indicate the Curie-Weiss
behavior without taking account of the FM fluctuations.~\cite{SCRFerro}
The large value of the $\mu _\text{p}$/$\mu _\text{s}$ also suggests the weak FM properties in this compound.

We consider two known examples of type-(ii) QCP for the 4\emph{f} electron system.
The well-known type-(ii) QCP doped compound is CeCu$_\text{5.9}$Au$_\text{0.1}$,
which exhibits the critical scaling behavior for the differential susceptibility ($\partial M/\partial H$) in the form
: $(\partial M/\partial H)^{-1}\!=\!\chi _{0} ^{-1}\!+CT^{\alpha} g(H/T)$.
Here $\alpha$ is the critical exponent and the universal scaling function $g(H/T)$
is given by Schr\"{o}der \emph{et al}..~\cite{CeCu6Au}
This scaling function, however, does not lead to the peak for $\partial M/\partial H$ ;
for this reason, it does not explain our results.
On the other hand, our results are very similar to the case of undoped YbRh$_2$Si$_2$,
which is classified as a type-(ii) QCP compound.
A plateau in the Knight shift below 1 K is observed in YbRh$_2$Si$_2$, while the ac susceptibility shows
a peak at the marginal temperature with the magnetic field dependence.~\cite{YbRhSi1,YbRhSi2,YbRhSi3}
The Curie-Weiss behavior of magnetic susceptibility $\chi(T)$ for YbRh$_2$Si$_2$, however, hints to large
fluctuating localized Yb$^{3+}$ moments, while that for our system indicates very tiny fluctuating moments
with itinerant nature.

We speculate that the magnetic properties of CeRu$_2$Si$_2$ at small magnetic fields and ultra low temperatures
are determined by competition between FM and AFM fluctuations.
The narrow range of the applied magnetic fields above 0.20 mT can modulate the FM fluctuations and show
a magnetic field arranged state below $T_\text{P}$.
The FM fluctuations, however, decrease with increasing magnetic field because the magnetic ground state is recovered
nearly to the LFL state at the field above 6.21 mT.
In the field of 0.02 mT, the FM and AFM fluctuations compete strongly and show a non-analytical
temperature dependence.
This possible scenario of the QCP in our compound shows that CeRu$_2$Si$_2$ at small magnetic fields
and ultra low temperatures can probably be considered one of the candidates for investigations of quantum phase transitions
at ambient pressure and without any doping.
Similar to the type-(ii) QCP compound YbRh$_2$Si$_2$, the external magnetic field is the control parameter for that transition.
The NMR measurements for CeRu$_2$Si$_2$ under the conditions described above would be very
useful in identifying of the nature of the itinerant 4\emph{f} electron system and that QCP.

This work was partially supported by a Grant-in-Aid for Scientific Research from the Ministry of
Education, Culture, Sports, Science and Technology of Japan. The first author is grateful to JSPS for a
Research Fellowship for Young Scientists. The authors are also grateful to H. Tsujii and Y. Tabata for
valuable discussions and to K. Mukai, T. Tsunekawa and K. Nunomura for technical support.


\begin{thebibliography}{99}
\bibitem{Doniach} S. Doniach, Physica B \textbf{91}, 231 (1977)
\bibitem{QCP1} See \emph{e.g.} Proceedings of the International Conference on Strongly Correlated Electron Systems, Physica B \textbf{281}\&\textbf{282} (2000)
\bibitem{QCP2} S. Sachdev, \emph{Quantum Phase Transitions}, (Cambridge Univ. Press, Cambridge 1999).
\bibitem{Millis} A. J. Millis, Phys. Rev. B \textbf{48}, 7183 (1993)
\bibitem{SCRHF} T. Moriya and T. Takimoto, J. Phys. Soc. Jpn. \textbf{64}, 960 (1996)
\bibitem{Si} Q. Si, S. Rabello, K. Ingersent, and J. L. Smith, Nature (London) \textbf{804} (2001)
\bibitem{CeCu6Au} A. Schr\"{o}der, G. Aeppli, R. Coldea, M. Adams, O. Stockert, H. v. L\"{o}hneysen, E. Bucher, R. Ramazashvili,
and P. Coleman, Nature (London) \textbf{351} (2000) and references in there.
\bibitem{YbRhSi1} K. Ishida, K. Okamoto, Y. Kawasaki, Y. Kitaoka, O. Trovarelli, C. Geibel, and F. Steglich, Phys. Rev. Lett. \textbf{89},
107202 (2002)
\bibitem{Gamma} M. J. Besnus, J. P. Kappler, P. Lehmann, and A. Meyer, Solid State Commun. \textbf{55}, 779 (1985)
\bibitem{KondoTemp} L. P. Regnault, W. A. C. Erkelens, J. Rossat-Mignod, P. Lejay, and J. Flouquet, Phys. Rev. B
\textbf{38}, 4481 (1988)
\bibitem{meta1} P. Haen, J. Flouquet, F. Lapierre, P. Lejay, and G. Remenyi, J. Low. Temp. Phys. \textbf{67}, 391 (1987)
\bibitem{meta2} T. Sakakibara, T. Tayama, K. Matsuhira, H. Mitamura, H. Amitsuka, K. Maezawa, and Y. \={O}nuki, Phys. Rev. B
\textbf{51}, R12030 (1995)
\bibitem{meta3} H. Satoh and F. J. Ohkawa, Phys. Rev. B, \textbf{57}, 5891 (1998)
\bibitem{meta4} S. Raymond, D. Raoelison, S. Kambe, L. P. Regnault, B. F$\mathring{\text{a}}$k, R. Calemczuk, J. Flouquet, P. Haen, and P. Lejay, Physica B \textbf{259-261}, 48 (1999)
\bibitem{meta5} M. Sato, Y. Koike, S. Katano, N. Metoki, H. Kadowaki, and S. Kawarazaki, J. Phys. Soc. Jpn. \textbf{70} Suppl. A,
118 (2001)
\bibitem{AFM1} J. Rossat-Mignod, L. P. Regnault, J. L. Jacoud, C. Vettier, P. Lejay, J. Flouquet, E. Walker, D. Jaccard, and
A. Amato, J. Magn. Magn. Mater. \textbf{76}\&\textbf{77}, 376 (1988)
\bibitem{AFM2} M. Sato, S. Kawarazaki, Y. Miyako, and H. Kadowaki, J. Phys. Chem. Solids \textbf{60}, 1203 (1999)
\bibitem{mSR} A. Amato, R. Feyerherm, F. N. Gygax, A. Schenck, J. Flouquet, and P. Lejay, Phys. Rev. B \textbf{50}, R619 (1994)
\bibitem{La1} H. Haen, J. P. Kappler, F. Lapierre, P. Lehmann, P. Lejay, J. Flouquet, and A. Meyer, J. Phys. C \textbf{8}, 757 (1988)
\bibitem{Rh1} Y. Miyako, T. Takeuchi, T. Taniguchi, S. Kawarazaki, K. Marumoto, R. Hamada, Y. Yamamoto, M. Ocio, P. Pari,
and J. Hammann, J. Phys. Soc. Jpn. \textbf{65} Suppl. B, 12 (1996)
\bibitem{MeltingCurve} D. S. Greywall, Phys. Rev. B \textbf{33}, 7520 (1986)
\bibitem{CeCu6static} E. A. Schuberth, J. Schupp, R. Freese, and K. Andres, Phys. Rev. B \textbf{51}, R12892 (1995)
\bibitem{S-K.Ma} S-K. Ma, \emph{Moden Theory of Critical Phenomena}, (W. A. Benjamin, 1976).
\bibitem{MFSDW} P. A. Fedders and P. C. Martin, Phys. Rev. \textbf{143}, 245 (1966)
\bibitem{Tsujii} H. Tsujii, E. Tanaka, Y. Ode, T. Katoh, T. Mamiya, S. Araki, R. Settai, and Y. \={O}nuki, Phys. Rev. Lett. \textbf{84},
5407 (2000) and references in there.
\bibitem{CharacteristicTemp} S. Kambe, J. Flouquet, and T. E. Hargreaves, J. Low Temp. Phys. \textbf{108}, 383 (1997)
\bibitem{YbRhSi2} O. Trovarelli, C. Geibel, S. Mederle, C. Langhammer, F. M. Grosche, P. Gegenwart, M. Lang, G. Sparn, and F. Steglich,
Phys. Rev. Lett. \textbf{85}, 626 (2000)
\bibitem{YbRhSi3} P. Gegenwart, J. Custers, C. Geibel, K. Neumaier, T. Tayama, K. Tenya, O. Trovarelli, and F. Steglich,
Phys. Rev. Lett. \textbf{89}, 056402 (2002)
\bibitem{SCRFerro} K. Usami and T. Moriya, J. Phys. Soc. Jpn. \textbf{44}, 122 (1978)
\end{thebibliography}

\end{document}